\documentclass[aps,preprint,amsmath,showpacs,showkeys]{revtex4}

\usepackage{bm}

\begin{document}

\title{Lepton flavor mixing in the Wolfenstein scheme}

\author{V.~Gupta}

\email[]{virendra@mda.cinvestav.mx}

\affiliation{
Departamento de F\'{\i}sica Aplicada.\\
Centro de Investigaci\'on y de Estudios Avanzados del IPN.\\
Unidad M\'erida.\\
A.P. 73, Cordemex.\\
M\'erida, Yucat\'an, 97310. MEXICO.
}

\author{G.~S\'anchez-Col\'on}

\email[]{gsanchez@mda.cinvestav.mx}

\affiliation{
Departamento de F\'{\i}sica Aplicada.\\
Centro de Investigaci\'on y de Estudios Avanzados del IPN.\\
Unidad M\'erida.\\
A.P. 73, Cordemex.\\
M\'erida, Yucat\'an, 97310. MEXICO.
}

\author{S.~Rajpoot}

\email[]{subhash.rajpoot@csulb.edu}

\affiliation{
Department of Physics \& Astronomy.\\
California State University, Long Beach.\\
Long Beach, CA 90840. USA.
}

\author{H-C.~Wang}

\email[]{Hsiching.Wang@csulb.edu}

\affiliation{
Department of Physics \& Astronomy.\\
California State University, Long Beach.\\
Long Beach, CA 90840. USA.
}

\date{\today}

\begin{abstract}

We mimic the Wolfenstein scheme of quark flavor mixing to describe neutrino
oscillations in the standard model. We identify the parameter $\lambda$ with
the experimentally measured value of the mixing element responsible for
atmospheric neutrino oscillations. The matrix elements responsible for solar
neutrino oscillations and the Chooz angle are taken to be proportional to
$\lambda^2$ and $\lambda^3$, respectively. Using present world average data on
neutrinos, we derive bounds on the other parameters $A$, $\rho$, and $\eta$,
of the new scheme.

\end{abstract}


\maketitle

Activities in neutrino physics over the past several years have accumulated
substantial data to establish a near robust picture of neutrino properties. All
experiments confirm that neutrinos have {\it tiny} masses and  {\it oscillate}.
The appropriate squared mass differences $\Delta m^2_{ij}$ and the mixing
angles $\theta_{ij}$ $(i,j=1,2,3; i<j)$ are probed for solar
neutrinos~\cite{btc,sage,gallex,sno,sk}, atmospheric
neutrinos~\cite{sk2,soudan}, reactor neutrinos~\cite{kl1,boehm,chooz}, and
accelerator neutrinos~\cite{k2k,minosold,t2k,minos,doublechooz,dayabay}. At
present, empirical information on whether neutrinos are massive Dirac or
Majorana particles is still lacking. Also, the absolute values of neutrino
masses are unknown. Although the complete picture seems a long way in the
future, enough information exists to create a neutrino paradigm on the broader
aspects of neutrino properties. Hopefully, to complete the picture, future
experiments will unravel more information on neutrinos pertaining to their
absolute masses, the phenomenon of neutrino CP violation and any symmetries,
discrete or continuous, lurking in lepton flavor mixing. In the following we
will assume that there are only three neutrino species.

Neutrinos ($\nu_e$, $\nu_{\mu}$, $\nu_{\tau}$), emitted in weak decays,
are admixtures of different mass eigenstates ($\nu_1$, $\nu_2$, $\nu_3$).
This leads to leptonic flavors mixing that is responsible for the presently
observed neutrino oscillations. The lepton flavor mixings originate in the
charged current weak interactions Lagrangian ${\cal L}_{\rm cc}$. It is
described by the $3\times 3$ unitary matrix $U$, the PMNS mixing
matrix~\cite{pont,mns}, analogous to the CKM mixing matrix for
quarks~\cite{cab,km}:

\begin{eqnarray}
-{\cal L}_{\rm cc} = \frac{g}{\sqrt{2}}
\left(\begin{array}{ccc} \overline{e} & \overline{\mu} & \overline{\tau}
\end{array}\right)_{\rm L} \, \gamma^\mu \left(\begin{array}{ccc} U^{}_{e1}
& U^{}_{e2} & U^{}_{e3} \\ U^{}_{\mu 1} & U^{}_{\mu 2} & U^{}_{\mu 3} \\
U^{}_{\tau 1} & U^{}_{\tau 2} & U^{}_{\tau 3} \end{array} \right)
\left(\begin{array}{c} \nu^{}_1 \\ \nu^{}_2 \\ \nu^{}_3 \end{array}
\right)_{\rm L} W^-_\mu + {\rm h.c.}
\end{eqnarray}

\noindent
The neutrino flavor eigenstates are related to the corresponding mass
eigenstates as

\begin{equation}
\left(\begin{array}{l} \nu _{e} \\ \nu
_{\mu } \\ \nu _{\tau } \end{array} \right)_{\rm L} =U\left( \begin{array}{l} \nu
_{1} \\ \nu _{2} \\ \nu _{3} \end{array} \right)_{\rm L}
\end{equation}

The elements of the matrix $U$ can be complex and probe directly the mixing
between the three neutrinos. This parameterization will be relevant when we
discuss the Wolfenstein scheme for neutrinos. However, the standard
parameterization of $U$ is in terms of three angles ($\theta^{}_{12}$,
$\theta^{}_{23}$, $\theta^{}_{13}$) and three phases ($\delta$, $\rho$,
$\sigma$). Explicitly,

\begin{eqnarray}
U &=&R_{23}(\theta_{23})\,P^{\dagger}_{\delta}\,
   R_{13}(\theta_{13})\,P_{\delta}\,R_{12}(\theta_{12}) P_{\nu}  \\ \nonumber
& & \\ \nonumber
&=&\left(
\begin{array}{ccc}
1 & 0 & 0 \\
0 & c_{23} & s_{23} \\
0 & -s_{23} & c_{23}
\end{array}
\right) \left(
\begin{array}{ccc}
c_{13} & 0 & s_{13}e^{-i\delta } \\
0 & 1 & 0 \\
-s_{13}e^{i\delta } & 0 & c_{13}
\end{array}
\right) \left(
\begin{array}{ccc}
c_{12} & s_{12} & 1 \\
-s_{12} & c_{12} & 0 \\
0 & 0 & 1
\end{array}
\right) P_{\nu}\\ \nonumber
& & \\ \nonumber
&=&\left(
\begin{array}{ccc}
c_{13}c_{12} & c_{13}s_{12} & s_{13}e^{-i\delta } \\
-s_{12}c_{23}-c_{12}s_{23}s_{13}e^{i\delta } &
c_{12}c_{23}-s_{12}s_{23}s_{13}e^{i\delta } & s_{23}c_{13} \\
s_{12}s_{23}-c_{12}c_{23}s_{13}e^{i\delta} &
-c_{12}s_{23}-s_{12}c_{23}s_{13}e^{i\delta } & c_{23}c_{13}
\end{array}
\right)P_{\nu}, \\ \nonumber
\end{eqnarray}

\noindent
where $R_{jk}(\theta_{jk})$ describes a rotation in the $jk$-plane
through angle $\theta_{jk}$ and $c^{}_{ij} \equiv \cos\theta^{}_{ij}$,
$s^{}_{ij} \equiv \sin\theta^{}_{ij}$ (for $ij = 12, 13, 23$). The phase
matrices $P^{}_{\delta }$ and  $P^{}_{\nu }$ are defined as

\begin{equation}
P^{}_{\delta}=
\left(\begin{array}{ccc} 1 & 0 & 0 \\ 0 &1 &
0 \\ 0 & 0 & e^{-i\delta} \end{array} \right),
\quad
P^{}_{\nu}=
\left(\begin{array}{ccc} e^{i\rho} & 0 & 0 \\  0 & e^{i\sigma} & 0 \\ 0 & 0 &
1\end{array} \right),
\end{equation}

\noindent
where $P^{}_{\nu}$ is relevant only if the neutrinos are massive Majorana
particles while the Dirac phase $\delta $ is always present irrespective of the
nature of neutrinos. Since information on absolute masses for neutrinos is
lacking, the phase matrix $P_{\nu}$ will not be considered further.

In the standard parameterization of $U$, the oscillation parameters are the
mixing angles ($\theta_{12}$, $\theta_{23}$, $\theta_{13}$) and the mass
squared differences $\Delta m^2_{12}$ and $\Delta m^2_{23}$. The parameters
$\Delta m^2_{12}$ and $\theta_{12}$ describe solar
neutrinos~\cite{btc,sage,gallex,sk,sno} and the reactor experiment
KamLAND~\cite{kl1, kl2}; $\Delta m^2_{23}$ and $\theta_{23}$ describe
atmospheric neutrino oscillations~\cite{sk2,soudan} and the K2K accelerator
experiment~\cite{k2k}; while $\theta_{13}$ and $\Delta m^2_{13}$ describe
neutrino oscillations in the T2K, MINOS, Double Chooz, Daya Bay and RENO
experiments~\cite{k2k,minosold,t2k,minos,doublechooz,dayabay,reno}. For
normal neutrino mass hierarchy ($m_3 \ge m_2 \ge  m_1$) the sum total of the
present knowledge on these parameters, averaged over all experiments, at
various statistical significance levels is collected in Table~\ref{table1}.

Neutrinoless double beta decay and cosmology provide direct information on the
mass parameters that are complementary to the oscillation parameters. The
tritium experiments~\cite{tri} provide an upper bound on the absolute value
of neutrino mass

\begin{equation}
m_i \leq 2.2\,\rm{eV}.
\end{equation}

\noindent
A more strict bound

\begin{equation}
m_i \leq 0.6\,\rm{eV},
\end{equation}

\noindent
follows from the analysis of the cosmological data~\cite{cos}.

The data in Table~\ref{table1} indicates that neutrino oscillations are
described by two large mixing angles: $\theta^{}_{12} \simeq 34^\circ$ and
$\theta^{}_{23} \simeq 45^\circ$. The third mixing angle $\theta^{}_{13}$ is
much smaller. The results of the experiments T2K~\cite{t2k},
MINOS~\cite{minos}, Double Chooz~\cite{doublechooz}, DAYA BAY~\cite{dayabay},
and RENO~\cite{reno}, imply that this angle is non-zero and about $9^\circ$.

In our analysis, for normal neutrino mass hierarchy, we will use the neutrino
oscillation parameters spanning over the $3\sigma$ range~\cite{Fogli}:

\begin{eqnarray}
0.259 \leq &\sin^2 \theta^{}_{12}& \leq 0.359, \nonumber\\
0.331 \leq &\sin^2 \theta^{}_{23}& \leq 0.637,\\
0.017 \leq &\sin^2 \theta^{}_{13}& \leq 0.031, \nonumber
\end{eqnarray}

\noindent
and

\begin{eqnarray}
6.99\times 10^{-5}~{\rm eV}^2 \leq &\Delta m^2_{12}& \leq 8.18 \times
10^{-5}~{\rm eV}^2, \\
2.19\times 10^{-3}~{\rm eV}^2 \leq & \Delta m^2_{23}
& \leq 2.62 \times 10^{-3}~{\rm eV}^2. \nonumber
\end{eqnarray}

\noindent
One can derive the following $3\sigma$ CL ranges on the magnitude of the
elements of the leptonic mixing matrix $U$,

\begin{equation}
\label{eq:upmns}
  |U| = \left(
\begin{array}{ccc}
    0.795 \to 0.846 &
   \quad 0.513 \to 0.585 &
   \quad  0.126 \to 0.178
    \\
    0.205 \to 0.543 &
    \quad 0.416 \to 0.730 &
    \quad 0.579 \to 0.808
    \\
    0.215 \to 0.548 &
    \quad 0.409 \to 0.725 &
    \quad 0.567 \to 0.800
\end{array}
\right).
\end{equation}

\noindent
The ranges in the different entries of the matrix $U$ are correlated due to the
constraints imposed by unitarity.

In 1983, Wolfenstein~\cite{wolf} introduced an elegant scheme for describing the
quark mixing~\cite{cab,km} matrix $V_{\rm CKM}$. In this scheme the three angles
($\theta_{12}$, $\theta_{23}$, $\theta_{31}$) and the CP-violating phase
$\delta$ of $V_{\rm CKM}$ are replaced by four new parameters ($\lambda$, $A$,
$\rho$, $\eta$). The parameter $\lambda = \sin\theta_C=0.22$ represents the
Cabibbo angle, a well determined quantity experimentally, $A$ represents the
\lq\lq sizing\rq\rq parameter and $\rho$ and $\eta$ are CP violating
parameters. In terms of the new parameters $V_{\rm CKM}$ takes an elegant form,

\begin{equation}
V_{\rm CKM}=
\left(
\begin{array}{ccc}
1-\frac{1}{2}\lambda^2 & \lambda &
A\lambda^3(\rho - i\eta) \\
-\lambda & 1-\frac{1}{2}\lambda^2 &
A\lambda^2\\
A\lambda^3(1-\rho-i\eta) & -A\lambda^2 & 1 \end{array}
\right).
\end{equation}

\noindent
This matrix gives a good description of data on quark mixing. In Wolfenstein's
parameterization the parameter $\lambda$ is identified with the experimentally
best known angle, the Cabibbo angle, and the other elements are identified with
quantities of order $\lambda^2$ and $\lambda^3$.

In the case of neutrinos~\cite{pont,mns} we adopt the same philosophy. We
identify the parameter $\lambda$ with the element $U_{\mu3}$ describing
atmospheric neutrino oscillations. The matrix element $U_{e2}$ describing
solar neutrino oscillations is assigned order $\lambda^2$ and the matrix
element $U_{e3}$ is assigned order $\lambda^3$ which gets identified with the
mixing Chooz angle $\theta_{13}$. Explicitly,

\begin{eqnarray}
U_{\mu3}&=& \lambda, \nonumber\\
U_{e2}&=&A\lambda^2,\\
U_{e3}&=&A\lambda^3(\rho-i\eta). \nonumber
\end{eqnarray}

\noindent
It follows that,

\begin{eqnarray}
\sin \theta_{13}&=&A\lambda^3 \sqrt{\rho^2+\eta^2}, \nonumber\\
\sin \theta_{12}&=&\frac{A\lambda^2}{\sqrt{1-A^2\lambda^6(\rho^2+\eta^2)}},\\
\sin \theta_{23}&=&\frac{\lambda}{\sqrt{1-A^2\lambda^6(\rho^2+\eta^2)}},\nonumber\\
\tan\delta&=&\frac{\eta}{\rho}. \nonumber
\end{eqnarray}

\noindent
Unitarity of the PMNS matrix $U$ determines the remaining elements:

\begin{eqnarray}
U_{e1}&=&
\sqrt{1-A^2\lambda^4-A^2\lambda^6(\rho^2+\eta^2)} \nonumber \\
& & \nonumber \\
U_{\mu1}&=&\frac{-A\lambda^2\sqrt{1-\lambda^2-A^2\lambda^6(\rho^2+\eta^2)}\,
-A\lambda^4\sqrt{\rho^2+\eta^2}
\sqrt{1-A^2\lambda^4-A^2\lambda^6(\rho^2+\eta^2)}\, e^{i \delta}}
{1-A^2\lambda^6(\rho^2+\eta^2)} \nonumber \\
& & \nonumber \\
U_{\mu2}&=&\frac{\sqrt{1-A^2\lambda^4-A^2\lambda^6(\rho^2+\eta^2)}
\sqrt{1-\lambda^2-A^2\lambda^6(\rho^2+\eta^2)}
-A^2\lambda^6\sqrt{\rho^2+\eta^2}\, e^{i \delta}}
{1-A^2\lambda^6(\rho^2+\eta^2)} \\
& & \nonumber \\
U_{\tau1}&=&\frac{-A\lambda^3\sqrt{\rho^2+\eta^2}\,
\sqrt{1-A^2\lambda^4-A^2\lambda^6(\rho^2+\eta^2)}\,
\sqrt{1-\lambda^2-A^2\lambda^6(\rho^2+\eta^2)}\, e^{i \delta}+A\lambda^3}
{1-A^2\lambda^6(\rho^2+\eta^2)} \nonumber\\
& & \nonumber \\
U_{\tau2}&=&\frac{-A^2\lambda^5\sqrt{\rho^2+\eta^2}\,
\sqrt{1-\lambda^2-A^2\lambda^6(\rho^2+\eta^2)}\, e^{i\delta}
-\lambda\sqrt{1-A^2\lambda^4-A^2\lambda^6(\rho^2+\eta^2)}}
{1-A^2\lambda^6(\rho^2+\eta^2)} \nonumber \\
& & \nonumber \\
U_{\tau3}&=&\sqrt{1-\lambda^2-A^2\lambda^6(\rho^2+\eta^2)}.
\end{eqnarray}

\noindent
Notice that, unlike the case of Wolfenstein scheme for quarks where
flavor mixing matrix elements are expanded in terms of the parameter $\lambda$,
we provide exact results for the matrix elements of the PMNS matrix $U$ in
terms of our parameter $\lambda$ for neutrinos.

Next, we determine the constraints on the remaining parameters $A$, $\rho$, and
$\eta$ from present data on neutrino oscillations. We take, from
Eq.~(\ref{eq:upmns}), the following range of values for $U_{\mu3}$, $U_{e2}$,
and $U_{e3}$, to stretch over the $3\sigma$ CL:

\begin{eqnarray}
0.808\geq& U_{\mu3}& \geq 0.579, \nonumber\\
0.585\geq& U_{e2}& \geq 0.513,\\
0.178 \geq& U_{e3}&  \geq 0.126; \nonumber
\end{eqnarray}

\noindent
with this as input, we determine the following constraints on the \lq\lq
sizing\rq\rq parameter $A$ and the \lq\lq CP\rq\rq  parameters $\rho$ and
$\eta$:

\begin{equation}
1.530\geq A \geq 0.896,
\quad\quad
0.424\geq \sqrt{\rho^2+\eta^2} \geq 0.377.
\end{equation}

The Jarlskog~\cite{jarlskog} parameter $J$ that provides a measure of CP
violation is given by

\begin{equation}
J=
\frac{\sqrt{[1-\lambda^2-A^2\lambda^6(\rho^2+\eta^2)]
[1-A^2\lambda^4-A^2\lambda^6(\rho^2+\eta^2)]}}
{\sqrt{\rho^2+\eta^2}}\, A^2\lambda^6 \eta.
\end{equation}

\noindent
Maximal CP violation corresponds to $\rho = 0$ while CP conservation
corresponds to $\eta=0$. Presently, extracting the CP-violating phase
$\delta$ from neutrino experiment is a challenging task. However, the situation
looks promising because $\delta$ is tied to the mixing angle $\theta_{13}$
which itself is as large as the Cabibbo angle. Additionally, even if
$\rho^2+\eta^2$ is accessed phenomenologically, there will still remain the
ambiguity of knowing to which of the four quadrants $\delta$ belongs.

\begin{acknowledgments}

V.~Gupta and G.~S\'anchez-Col\'on would like to thank CONACyT (M\'exico) for
partial support. The work of S. Rajpoot was supported by DOE Grant No:
DE-FG02-10ER41693.

\end{acknowledgments}

\clearpage

\begin{table}

\caption{The latest global-fit results of three neutrino mixing
angles $(\theta^{}_{12}, \theta^{}_{23}, \theta^{}_{13})$ and two
neutrino mass-squared differences $\Delta m^2_{12} \equiv | m^2_2 - m^2_1|$
and $\Delta m^2 _{23}\equiv |m^2_3 - m^2_2|$ in the case of
normal neutrino mass hierarchy~\cite{Fogli}.\label{table1}}

\begin{ruledtabular}

\begin{tabular}{cccccc}

  & \multicolumn{5}{c}{Parameter} \\
  & $\Delta m^2_{12}$ & $\Delta m^2_{23}$
  & $\theta^{}_{12}$ & $\theta^{}_{23}$ & $\theta^{}_{13}$ \\
  & $(10^{-5}~{\rm eV}^2)$ & $(10^{-3}~{\rm eV}^2)$ & & & \\

\hline

  Best fit & $7.54$ & $2.43$ & $33.6^\circ$ & $38.4^\circ$ & $8.9^\circ$ \\
  $1\sigma$ range & $[7.32, 7.80]$ & $[2.33, 2.49]$ & $[32.6^\circ, 34.8^\circ]$
  & $[37.2^\circ, 40.0^\circ]$ & $[8.5^\circ, 9.4^\circ]$ \\
  $2\sigma$ range & $[7.15, 8.00]$ & $[2.27, 2.55]$ & $[31.6^\circ, 35.8^\circ]$
  & $[36.2^\circ, 42.0^\circ]$ & $[8.0^\circ, 9.8^\circ]$ \\
  $3\sigma$ range & $[6.99, 8.18]$ & $[2.19, 2.62]$ & $[30.6^\circ, 36.8^\circ]$
  & $[35.1^\circ, 53.0^\circ]$ & $[7.5^\circ, 10.2^\circ]$ \\

\end{tabular}

\end{ruledtabular}

\end{table}

\end{document}